\documentclass[12pt, twoside]{article}

\usepackage{latexsym}
\usepackage{epsfig,graphics}
\usepackage{amsmath,amssymb,amsthm}
\usepackage{graphics,epsfig,calc}
\usepackage{upref}
\usepackage{color}
 \newcommand{\beqn}{\begin{eqnarray}}
 \newcommand{\eeqn}{\end{eqnarray}}
 \newcommand{\ba}{\begin{array}}
 \newcommand{\ea}{\end{array}}
\newcommand{\bc}{\begin{cor}}
 \newcommand{\ec}{\end{cor}}

 \newcommand{\pa}{\partial}
 \newcommand{\ds}{\displaystyle}
 \newcommand{\rIm}{{\rm Im\5}}

\newcommand{\fr}{\frac}

\newcommand{\ov}{\overline}

\newcommand{\dv}{{\rm div~}}

\newcommand{\cH}{{\cal H}}

\newcommand{\cL}{{\cal L}}

\newcommand{\cM}{{\cal M}}

\newcommand{\cO}{{\cal O}}

\newcommand{\ve}{\varepsilon}
\newcommand{\vp}{\varphi}
\newcommand{\De}{\Delta}
\newcommand{\de}{\delta}

\newcommand{\al}{\alpha}

\newcommand{\si}{\sigma}

\newcommand{\na}{\nabla}

\newcommand{\5}{{\hspace{0.5mm}}}

\newcommand{\R}{\mathbb{R}}

\newcommand{\C}{\mathbb{C}}

\newcommand{\nc}{\newcommand}
\nc{\dps}{\displaystyle}

\nc{\RR}{\mbox{\rm I$\!$R}}

\newcommand{\bd}{\begin{defin}}
 \newcommand{\ed}{\end{defin}}
\newcommand{\bt}{\begin{theorem}}
 \newcommand{\et}{\end{theorem}}
\newcommand{\bqt}{\begin{qtheorem}}
 \newcommand{\eqt}{\end{qtheorem}}

\newcommand{\bl}{\begin{lemma}}
 \newcommand{\el}{\end{lemma}}

 \newcommand{\bce}{\begin{center}}
 \newcommand{\ece}{\end{center}}

\newcommand{\bex}{\begin{example}}
 \newcommand{\eex}{\end{example}}
\newcommand{\bexs}{\begin{examples}}
 \newcommand{\eexs}{\end{examples}}

\newcommand{\bexe}{\begin{exercice}}
 \newcommand{\eexe}{\end{exercice}}

\newcommand{\brs}{\begin{remarks}}
 \newcommand{\ers}{\end{remarks}}

 \newcommand{\bpr}{\begin{proof}}
\newcommand{\epr}{\end{proof}}

\newtheorem{theorem}{Theorem}[section]
\newtheorem{qtheorem}{QTheorem}[section]

\newtheorem{defin}[theorem]{Definition}

\newtheorem{lemma}[theorem]{Lemma}
\newtheorem{remark}[theorem]{Remark}
\newtheorem{remarks}[theorem]{Remarks}
\newtheorem{cor}[theorem]{Corollary}
\newtheorem{pro}[theorem]{Proposition}
\newcommand{\bp}{\begin{pro}}
\newcommand{\ep}{\end{pro}}

\begin{document}
\begin{titlepage}
\bigskip\bigskip\bigskip

\begin{center}
{\Large\bf
On Dynamical Justification of Quantum
\bigskip\\
 Scattering Cross Section
}
\vspace{1cm}
 \bigskip\bigskip\\
{\large Alexander Komech}
\footnote{
Supported partly
by Alexander von Humboldt Research Award,
Austrian Science Fund (FWF): P22198-N13,
and the grants of
DFG and RFBR.}\\
{\it Fakult\"at f\"ur Mathematik, Universit\"at Wien\\
and Institute for Information Transmission Problems RAS
}\\
 e-mail:~alexander.komech@univie.ac.at

\end{center}

\date{}

\vspace{1cm}
\begin{abstract}

A~dynamical justification of quantum differential cross section
in the context of long time transition to stationary regime for the Schr\"odinger equation is suggested.
The problem has been stated by Reed and Simon.

Our approach is based on spherical incident waves produced by a harmonic source
and the long-range asymptotics for the corresponding spherical
limiting amplitudes. The main results are as follows:
i)~the convergence of  spherical limiting amplitudes to the limit as the source increases to infinity, and
ii) the universally recognized 
formula for the differential cross section corresponding to the
limiting flux.

The main technical ingredients are the Agmon--Jensen--Kato's  analytical theory of the Green function,
Ikebe's uniqueness theorem for the Lippmann--Schwinger equation, and
some adjustments of classical asymptotics for the Coulomb potentials.

\end{abstract}

\medskip


\end{titlepage}


\setcounter{equation}{0}
\setcounter{theorem}{0}
\section{Introduction}
\begin{figure}[!ht]
\vspace{0.5cm}
\centering
\hspace{-2cm}
\includegraphics[width=0.8\textwidth]{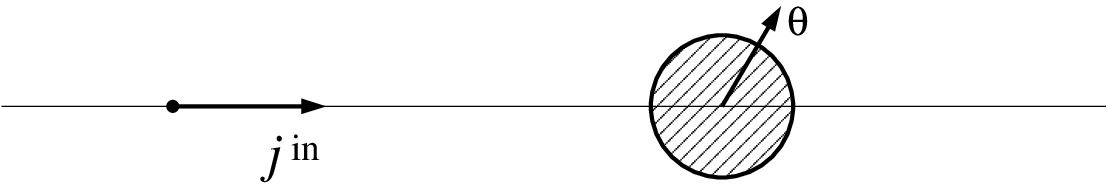}
\caption{Incident flux and scatterer.}
\label{Picture14}
\end{figure}

The differential  cross section is the main observable in quantum
scattering
experiments. This concept was first introduced to describe the Rayleigh scattering of sunlight and the Rutherford alpha-particle scattering as
the quotient
\begin{equation}\label{scs}
  \si(\theta)={j_a^{\rm sc}(\theta)}/|{j^{\rm in}}|.
\end{equation}
Here,  $j^{\rm in}$ is the incident stationary flux, and
$j_a^{\rm sc}(\theta)$ is  the angular density of
the scattered
stationary flux
$j^{\rm sc}(x)$ in the direction $\theta\in\R^3$, $|\theta|=1$ (see Fig.~\ref{Picture14}):
\begin{equation}\label{jout}
  {j_a^{\rm sc}(\theta)}=\lim_{R\to\infty} R^2j^{\rm sc}(R\theta) \cdot\theta.
\end{equation}
In both scattering processes studied by Rayleigh and  Rutherford 
the concept of differential cross section is well-established in the framework
of the corresponding dynamical equations:
the Maxwell equations in the case of Rayleigh scattering and the Newton equations  in the case of Rutherford scattering.

On the other hand, a satisfactory dynamical justification
of quantum  scattering cross section is still  missing
in the framework of the Schr\"odinger equation
\begin{equation}\label{S2}
  i\dot\psi(x,t)=H\psi(x,t)
:=-\fr 12 \De\psi(x,t)+V(x)\psi(x,t), \qquad x\in\R^3.
\end{equation}
The problem has been stated and discussed by Reed and Simon
in \cite{RS},  pp. 355--357. We suggest the solution for the first time, as far as we are aware.
The corresponding charge  and flux densities are defined as
\begin{equation}\label{Sflu}
  \rho(x,t)=|\psi(x,t)|^2, \qquad j(x,t)=\rIm [\ov{\psi(x,t)}\na\psi(x,t)].
\end{equation}
These densities satisfy the charge continuity equation
\begin{equation}\label{cce}
  \dot\rho(x,t)+\dv j(x,t)=0, \qquad (x,t)\in\R^4.
\end{equation}
We justify the formula for the differential cross section
\begin{equation}\label{dics}
  \si(k,\theta)= 16\pi^4|T(|k|\theta,k)|^2,
\end{equation}
which is universally recognized in physical and mathematical
literature (see, for example, \cite{ES,New,RS,Ta,Weinberg}).
We denote by $k\in\R^3$  the `wave vector' of  the
incident plane wave
\begin{equation}\label{ipw}
  \psi^{\rm in}(x,t)=e^{i(kx-E_k t)}, \qquad E_k:=\fr 12k^2.
\end{equation}
Let the brackets $(\cdot,\cdot)$ denote the Hermitian scalar product in
the complex Hilbert space
$\cL^2:=L^2(\R^3)$, as well as its extension
to the duality between the weighted Agmon--Sobolev spaces, see (\ref{L2si}) and (\ref{7}).
The
$T${\it -matrix} is given by
\begin{equation}\label{Tm}
  T(k',k):=\fr 1{(2\pi)^3}(T(E_k+i0)e^{ikx},e^{ik'x} ), \qquad k', k\in\R^3,
\end{equation}
which is the integral kernel of the operator
$T(E_k+i0):=V-VR(E_k+i0)V$ (see Section 25 of \cite{KK2012})
in  the Fourier transform
\begin{equation}\label{Ft}
\hat\psi(k)=\int e^{-ikx}\psi(x)dx, \qquad \psi\in C_0^\infty(\R^3).
\end{equation}
Here,  $R(E):=(H-E)^{-1}$ is the resolvent of the Schr\"odinger operator $H$.

It is well known that the integral kernel $S(k',k)$ of the
scattering operator $S$ in the Fourier transform reads as
\begin{equation}\label{SF}
S(k',k)=\de(k'-k)-i\pi\de(E_{k'}-E_k)T(k',k),~~~k',k\in\R^3
\end{equation}
(see \cite{BS,New,RS,Ta}).
The commonly used  `naive scattering theory'
consists of the following statements \cite{RS,Schiff, Ta}.
\medskip\\
{\bf I.} The incident wave is identified with the plane wave (\ref{ipw}), which
propagates in the direction of the wave vector $k$
and is a solution to the free Schr\"odinger equation (\ref{S2})
with $V(x)=0$.
\begin{figure}[!ht]
\vspace{0.5cm}
\centering
\hspace{-2cm}
\includegraphics[width=0.7\textwidth]{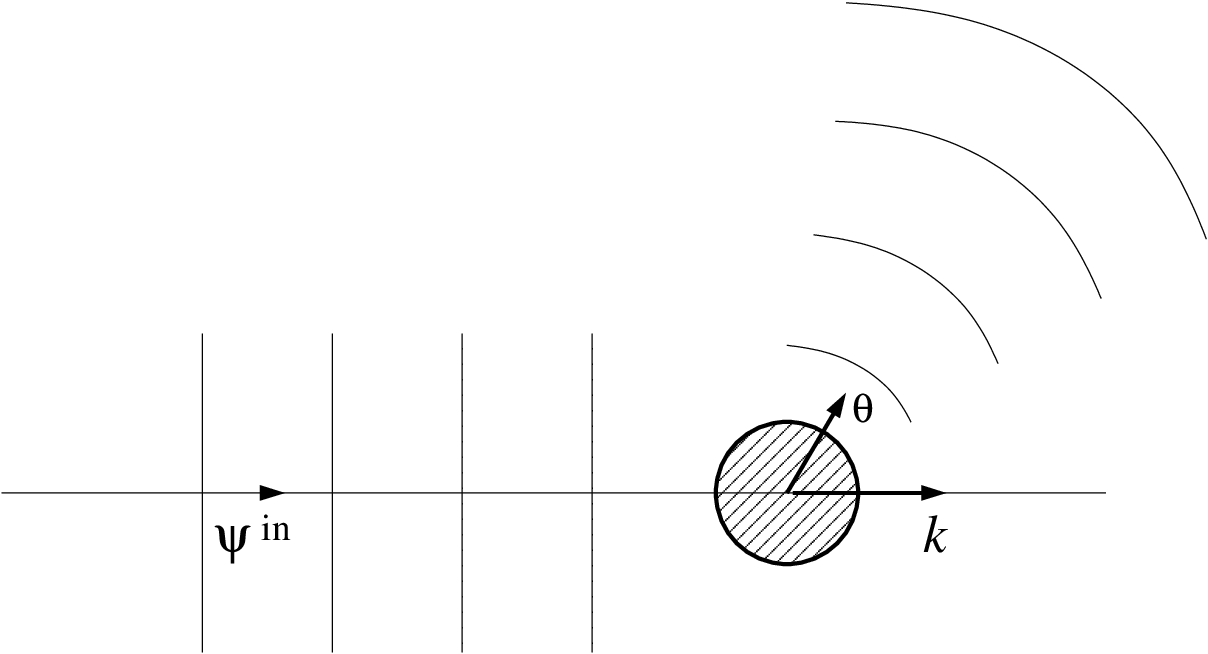}
\caption{Incident plane wave  and  outgoing spherical wave.}
\label{Picture15}
\end{figure}
\\
{\bf II.} The corresponding `scattered' solution to (\ref{S2})
is identified by its long time asymptotics on any bounded region $|x|<R$,
\begin{equation}\label{ltas}
  \psi(x,t)\sim A(x) e^{-iE_k t}, \qquad t\to\infty,
\end{equation}
where the amplitude  $A(x)$ is expressed by
\begin{equation}\label{ltast}
  A(x)=e^{ikx}-R(E_k+i0)[V(x)e^{ikx}].
\end{equation}
This amplitude admits the following
`spherical' long range asymptotics (3.58) of \cite[Ch.~4]{BS}
\begin{equation}\label{ecs}
  A(x)\sim e^{ikx}+a(k,\theta)\fr{e^{i|k|\cdot|x|}}{|x|},
  \qquad x|\to\infty, \qquad \theta:=x/|x|~;
\end{equation}
see Fig.~\ref{Picture15}.
\medskip\\
{\bf III.}
By (\ref{Sflu}), asymptotics (\ref{ecs}) give
\begin{equation}\label{jj}
  j^{\rm in}=k, \qquad j_a^{\rm sc}(\theta)=|a(k,\theta)|^2 |k|,
\end{equation}
and hence
the differential cross section reads
\begin{equation}\label{scsa}
  \si(k, \theta)=|a(k, \theta)|^2.
\end{equation}
It is well known   that  $a(k,\theta)$
is proportional to the $T$-matrix (formula (97a) of \cite{RS}):
\begin{equation}\label{aT}
  a(k,\theta)=-4\pi^2 T(|k|\theta,k).
\end{equation}
Hence,  (\ref{scsa}) reads as  (\ref{dics}).
\medskip

A {\it heuristic derivation}
of  relations (\ref{ltas}), (\ref{ltast})
 can be found in \cite{RS}, pp. 355--357.
However,
a mathematically consistent  justification
of the relations in a {\it time dependent picture}
was not suggested until now.
Moreover, relation
(\ref{dics})
was
considered up to now
as the {\it definition} of the differential cross section: see formulas (1.2) and (A.1.6) of \cite{ES}, formula (96) of
 \cite{RS}, and  Definition 7.9 on p.~254 of \cite{Yaf}.

The main problem in  mathematical justification of
(\ref{ltas}) and (\ref{ltast})
is related to the lack of a~consistent model for  the incident wave $\psi^{\rm in}(x,t)$,
securing convergence (\ref{ltas})
to a~stationary regime, and at the same time satisfies the
{\it `adiabatic condition'}
\begin{equation}\label{ad}
\psi^{\rm in}(x,t)\to 0, \qquad t\to-\infty,\quad x\in\R^3,
\end{equation}
which is in the spirit of the scattering theory.
The plane incident wave (\ref{ipw})
in the `naive scattering theory'
does not satisfy (\ref{ad}), since the wave occupies  the entire space.
The plane wave is a solution to the free Schr\"odinger equation
\begin{equation}\label{S0}
i\dot\psi(x,t)=-\fr 12 \De\psi(x,t), \qquad x\in\R^3.
\end{equation}
The adiabatic condition (\ref{ad})
in {\it acoustic scattering}
is provided by the `semi-infinite' incident plane wave
$$
\psi^{\rm in}(x,t)=\Theta(|k|t-kx)e^{i(kx-|k|t)}
$$
 for
$t<0$,
where $\Theta$ is the Heaviside function.
This incident wave is a solution to  the acoustic equation
\begin{equation}\label{aceq}
\ddot \psi(x,t)=\De\psi(x,t),\qquad |x|>R,
\end{equation}
for $t<-R$ if the scatterer is located in the region $|x|\le R$.
The similar incident plane wave can be constructed for the
Maxwell equations, which makes apparent the meaning of
the differential cross section in the Rayleigh scattering.

On the other hand,
a similar semi-infinite incident plane wave
does not exist
in the case of the Schr\"odinger equation. Indeed, we may  fix $R\gg |k|D$ and
take the semi-infinite  plane wave
$$
\psi^{\rm in}(x)=\Theta(-R-kx)e^{ikx}
$$
as the initial condition at $t=0$. However,
the corresponding solution does not satisfy the
 adiabatic condition for $t\to-\infty$.
The problem is of great importance also
in the context of the quantum field theory, where the
incident and outgoing plane waves
play the fundamental role \cite{New,Sakurai,Schiff, Weinberg}.

In the traditional approach,
the incident wave is a specific initial field, which is a~solution to the corresponding free wave equation
in the entire space. On the other hand, in practice, the incident wave is
a beam of particles or light
produced by a macroscopic source and satisfies the
free wave equation only outside the source.
One could expect that, for a large time, the incident wave
near the scatterer will asymptotically be a~free plane wave
 if the source is `monochromatic' and its distance from the scatterer,  $D$, tends to infinity.
This model obviously corresponds  to {\it spherical incident waves},
which are standard devices in optical and acoustic scattering \cite{Bor}.
\medskip

We justify formula (\ref{dics})
in the following
steps:
\medskip\\
A. First, we prove {\it the limiting amplitude principle}
for the Schr\"odinger equation (\ref{S2}) with harmonic source; i.e., the long time convergence
to a~stationary
harmonic regime with a~`spherical limiting amplitude',
which does not depend on initial state.
\medskip\\
B. Second, we prove the convergence
of the spherical
limiting amplitudes to the plane limiting amplitude
when the source goes off to infinity: $D\to\infty$.
\medskip\\
C. We deduce from  A and B  that  relations
(\ref{ltas})--(\ref{ecs}) and (\ref{scsa}) hold  true
in this double limit: first, as $t\to\infty$, and then, as $D\to\infty$.
\medskip\\
D. Finally,
we establish  the second relation of (\ref{jj}) for the scattered flux
\begin{equation}\label{scfi}
j^{\rm sc}(x,t):=j_\infty(x,t)-j^{\rm in},
\end{equation}
where $j_\infty(x,t)$ is the double limit of the current (\ref{Sflu}), and  $j^{\rm in}:=\lim_{|x|\to\infty} j_\infty(x,t)$.
Now formula (\ref{dics}) follows from (\ref{scsa}) and (\ref{aT}).
\medskip

Our technical novelties are as follows. We prove the limiting amplitude principle~A,
developing the Agmon--Jensen--Kato's theory
of the resolvent of the Schr\"odinger operator \cite{A,JK,KK2012}.
The proof of  the convergence B
relies on a~novel application of
Ikebe's uniqueness theorem for the Lippmann--\allowbreak Schwinger equation  \cite{BS,Ik} and on uniform bounds for
the Coulomb potentials (\ref{Kq}), (\ref{bah}), (\ref{unibo}). These bounds are due to  novel
asymptotics for
the Coulomb potentials
(\ref{asBS}),  which are
regularized at the zero point (the corresponding bound  (3.51) of \cite[Ch.~4]{BS},
is correct only for  $|x|\ge \de>0$ due to the singularity of the main term in the  asymptotics
(3.50) of \cite[Ch.~4]{BS}).
Moreover, we adjust the estimates  for the remainders
in the long range asymptotics of the Coulomb potentials, which dates back  to Povzner and Ikebe \cite{Ik,Po} (see (\ref{Kqg}) and (\ref{Kqg1})).

Note that formula (\ref{Sflu})
describes the {\it wave flux} corresponding to the
one-particle Schr\"odinger equation (\ref{S2}).
Respectively, the {\it many-particle interpretation} of  the cross section (\ref{scs})
is not straightforward. However, it is worth noting that definition
 (\ref{scfi}) means the principle of superposition for the currents
corresponding to the many-particle scattering.
 Thus,  the validity of the second relation of (\ref{jj}),
 as proved in our final Theorem \ref{tM}, suggests the many-particle interpretation.

 We note that  Theorem \ref{tM} was not established in Chapter 9 of \cite{KK2012},
 which is the previous version of present paper. Our progress relies on the novel
 estimates (\ref{Kqg}) and  (\ref{Kqg1}). Moreover, here we consider  general case of the Schr\"odinger
 operator $H$ with nonempty discrete spectrum, in contrast to \cite{KK2012}, Ch.~9.
 Finally, here we adjust our basic assumptions and proofs.

\medskip

We make some comments   on the known arguments for formula (\ref{dics}).
The traditional physical approach \cite{Ta} is based on random
incident wave packets $\psi^{\rm in}(x,0)$, which are asymptotically proportional to the plane waves $e^{ik x}$:
\begin{equation}\label{limd}
|\hat\psi^{\rm in}(k',0)|^2\to \de(k'-k).
\end{equation}
The known mathematical justifications reside in Dollard's fundamental result \cite{D69}
on {\it  scattering into cones}.
This result
 is used
in  \cite{T} for a~clarifying
treatment of formula (\ref{dics}).
Namely,
the normalized angular distribution of a~finite
 charge,  scattered for  infinite time,
converges to the
{\it normalized
function} (\ref{dics})  in the limit (\ref{limd}).

Dollard's result was refined in \cite{CNS,JLN} and in Section 3-3 of
 \cite{AJS},
where
{\it the flux across the surface theorem}
is proved.
This result was later
developed
in \cite{DGMZ,DGTZ,DMP,TDM} and applied for justification of formula (\ref{dics}) in the context of
the {\it  Bohmian particle mechanics} and incident stationary
random processes
 constructed of
normalized wave packets  (\ref{dics})  in the limit (\ref{limd}).
For a~survey, see~\cite{DT}. It is worth noting that we do not exclude the discrete spectrum
of the Schr\"odinger operator $H$, in contrast to~\cite{DGMZ}.
\medskip

We point out that all the previous results give the same expression
(\ref{dics})
for the differential cross section, though these results were not concerned with the
long time transition to a~stationary regime.
\medskip

Our paper is organized as follows.  The main results are stated in~Section \ref{MR}. The limiting amplitude principle is established in Section \ref{LAP}.
In Section \ref{SW} we obtain long range asymptotics and uniform bounds
for the
spherical limiting amplitudes. Next, in Sections 5 and~6 we prove convergence~B and the corresponding convergence for the flux.
Finally, in Sections~7 and~\ref{LRA} we verify formulas  (\ref{aT}) and~(\ref{jj}), which justify (\ref{scsa}) and~(\ref{dics}).
 \medskip

{\bf Acknowledgments.} The author thanks E. Kopylova and H. Spohn for useful discussions and remarks.

\setcounter{theorem}{0}
\setcounter{equation}{0}
\section{Main results}\label{MR}
We consider the Schr\"odinger equation with harmonic source:
\begin{equation}\label{Scc}
  \left\{\ba{l}i\dot\psi(x,t)=H\psi(x,t)+\rho_q(x)\5 e^{-iE_k t},
  \qquad t>0 \medskip \\ \psi(x,0)=\psi^0(x) \ea \right| \qquad x\in\R^3.
\end{equation}
Here, $H=-\fr12 \De+V(x)$, $E_k=k^2/2$ for $k\in\R^3\setminus 0$,
and $\rho_q(x):=|q|\rho(x-q)$ is the form factor of the source.

We mean that our model suits the physics of  quantum  scattering.
Namely, the incident wave is produced by time periodic source, like a~heated cathode in an electron gun.
The source is  not at infinity, though its distance from the scatterer is sufficiently large.

The solution $\psi(x,t)$ describes the spherical waves produced by the source.
The spherical waves look like plane waves near the scatterer in the limit $|q|\to\infty$.
The source is
described by a~density factor $\rho_q(x)$, where
the factor $|q|$ is introduced  for a~suitable normalization,
see (\ref{spw}) below.
\medskip

The weighted Agmon--Sobolev spaces $\cH^s_\si=\cH^s_\si(\R^3)$,  $s,\si\in\R$, are defined as follows.
Let $\cL^2_\si=\cL^2_\si(\R^3)$ be the Hilbert space of measurable functions in~$\R^3$ with norm
\begin{equation}\label{L2si}
\Vert \psi\Vert_{\cL^2_\si}^2=\int\langle x\rangle^{2\si}|\psi(x)|^2dx, \qquad \langle x\rangle:=\sqrt{x^2+1}.
\end{equation}
\bd
$\cH^s_\si=\cH^s_\si(\R^3)$ denotes the Hilbert space of tempered distributions $\psi(x)$ with finite norm
\begin{equation}\label{Hss}
 \Vert\psi\Vert_{\cH^s_\si}
 :=\Vert\langle \na\rangle^s\psi\Vert_{\cL^2_\si}<\infty.
\end{equation}
\ed
We will assume the following conditions.
\medskip\\
H0. The initial state $\psi^0$ is a~function from
 the  space
$\cH^2_{\si_0}$ with some $\si_0>5/2$.
\medskip\\
H1.
 For some $\ve_1>0$,
\begin{equation}\label{Vcr}
  \sup\limits_{x\in\R^3}\langle x\rangle^{4+\ve_1}|\pa^\al\rho(x)| <\infty,\qquad |\al|\le 2.
\end{equation}
%
%
%
%
\medskip\\
H2. The following Wiener condition holds:
\begin{equation}\label{Wr}
\hat\rho(|k|\theta):=\int e^{i|k|\theta x}\rho(x)dx\ne 0,
\qquad \theta\in\R^3, \ \ |\theta|=1.
\end{equation}
H3. The potential $V(x)$ is a~real $C^2$-function satisfying the condition
\begin{equation}\label{Vc}
  \sup\limits_{x\in\R^3}\langle x\rangle^{5+\ve_2}|\pa^\al V(x)|
<\infty,  \qquad
  |\al|\le 2,
\end{equation}
with some $\ve_2>0$.
 \medskip

Finally, we introduce our key  spectral assumption. Denote
$$
\cM_\si:=\{\psi\in \cL^2_{-\si}: \psi+R_0(0)V\psi=0\},
$$
where $R_0(E):=(-\fr12\De-E)^{-1}$ is the free resolvent.
The space $\cM_\si=\cM$ does not depend on
$\si\in (1/2,(5+\ve_2)/2)$ by the arguments preceding Lemma~3.1 of \cite{JK}.
 \medskip\\
H4.
We assume:
\begin{equation}\label{SCa}
{\bf The\  Spectral\ Condition:}
~~~~~~~~~~~~~~~~~~~\cM=0.
~~~~~~~~~~~~~~~~~~~~~~~~~~~~
\end{equation}
This condition
holds for {\it generic} potentials, see the discussion
preceding Lemma 3.1 in \cite{JK}.
 \medskip

\begin{figure}[!ht]
\vspace{1cm}
\centering
\hspace{-1cm}
\includegraphics[width=0.7\textwidth]{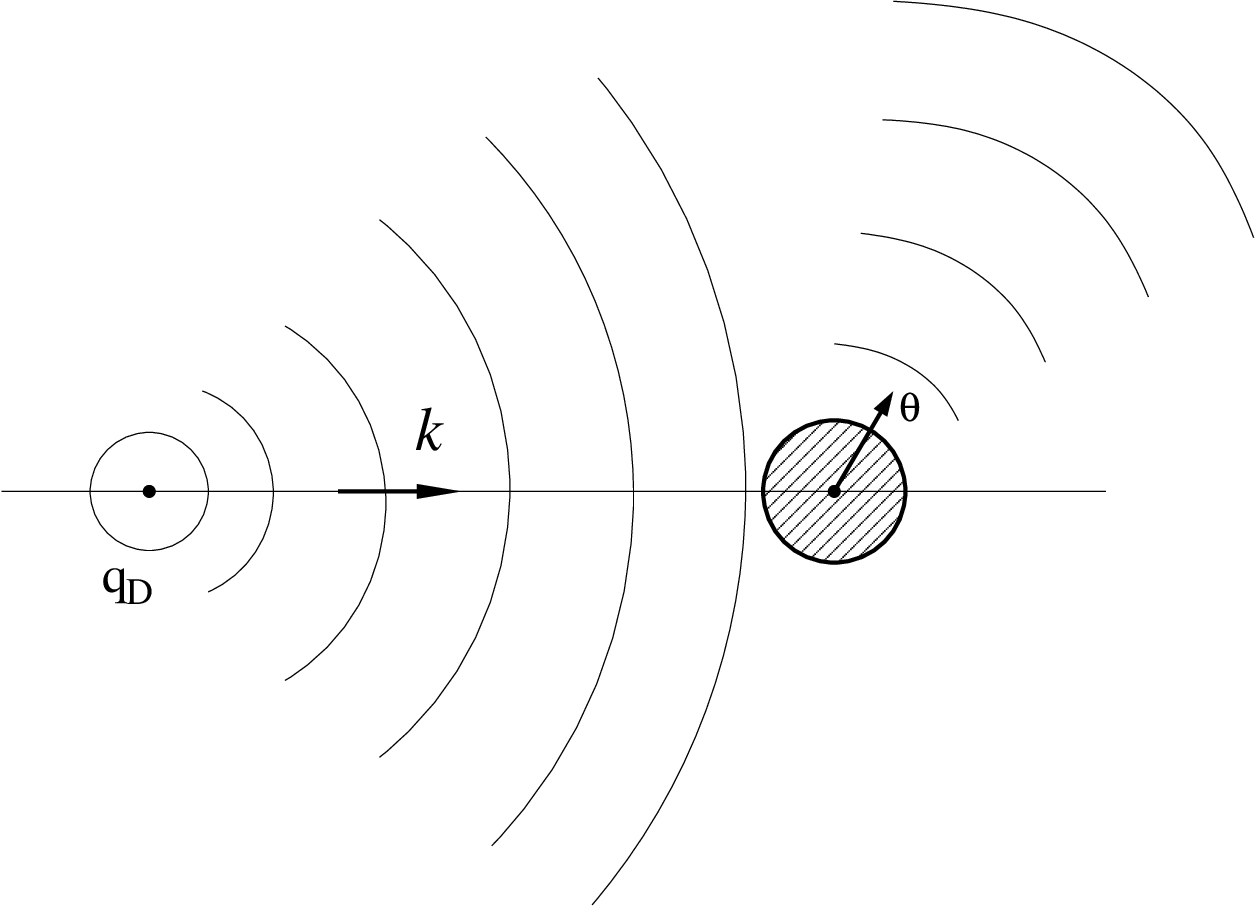}
\caption{Incident and outgoing spherical waves.}
\label{Picture16}
\end{figure}

Let us outline our plan.
\medskip\\
{\bf I.} First, we will prove the {\it limiting amplitude principle}:
\begin{equation}\label{laps}
~~
\psi(x,t)\sim
\vp_q(x,t)=
B_q(x)e^{-iE_k t}+\sum_1^NC^l_q\psi_l(x)e^{-iE^lt},~~~
t\to\infty,
\end{equation}
where $\psi_l(x)$ are the eigenfunctions of~$H$ corresponding to the eigenvalues $E^l<0$.
The asymptotics  hold in
 $\cH^2_{-\si}$
with  any
$\si>5/2$, and
the
{\it limiting amplitude} $B_q(x)$ is given by
\begin{equation}\label{las}
 B_q(x)=R(E_k+i0)\rho_q.
\end{equation}
The coefficients $C^l$ depend on the initial state
$\psi(x,0)$. On the other hand,
it is crucially important that
 the coefficients $C^l$ converge
as $|q|\to\infty$, while
the eigenfunctions $\psi_l(x)$ decay rapidly at infinity
by Agmon's theorem \cite{A}, Theorem 3.3  (see also Theorem 20.7 of \cite{KK2012}).
Hence,  the sum over the discrete spectrum
on the right-hand side of (\ref{laps})
does not contribute to the scattering cross section,
and we will omit it almost everywhere below.
\medskip\\
{\bf II.}
Second, denoting $B_D(x):=B_{q_D}(x)$, where
$q_D:=-nD$ with $n:=k/|k|$ and $D>0$, we establish
the following {\it `spherical version'} of  long range
asymptotics (\ref{ecs}):
\beqn\label{ecas}
~~~~~~~
B_D(x)&\sim&
b_D(n)
\Big[\fr{|q_D|}
{|x-q_D|} e^{i|k| (|x-q_D|-|q_D|)}
+
a_D(k,\theta)\fr{e^{i |k|\cdot|x|}}{|x|}\Big]
\nonumber\\
\nonumber\\
&&{\rm as}
\quad
|x-q_D|\to\infty,
\quad |x|\to\infty,
\eeqn
where $\theta:=x/|x|$
and
$b_D(n):=b(n)e^{i|k|D}$ with $b(n)\ne 0$; see Fig.~\ref{Picture16}.
The asymptotics  (\ref{ecas}) mean that the difference between the left-hand side and  the right-hand side
converges to zero.
\medskip\\
{\bf III.}
Further, we prove
the  convergence of the  spherical limiting
amplitudes, which is our central result: for $k\ne 0$
\begin{equation}\label{djR}
 A_D(x):=B_D(x)/b_D(n)
\to A(x), \qquad D\to\infty,
\end{equation}
where $A(x)$ is expressed by  (\ref{ltast}).
\medskip\\
{\bf IV.}
At last,  (\ref{djR}) implies the  asymptotics
 of the corresponding {\it limiting
solutions} $\vp_D(x,t):=\vp_{q_D}(x,t)$ (neglecting the last sum of (\ref{laps}))
\begin{equation}\label{laps2}
~~
\vp_D(x,t)/b_D(n)
\to
A(x)e^{-iE_k t},
\qquad D\to\infty, \quad (x,t)\in\R^4,
\end{equation}
and
of the corresponding
 flux (\ref{Sflu}):
\beqn\label{flu}
\!\!\!\!\!\!\!\!\!\!\!\!\!\!\!\!
\!\!\!\!
j_D(x)=\rIm [\ov{\vp_D(x,t)}\na \vp_D(x,t)]\!\!\!\!
&\!\!\!\!\longrightarrow\!\!\!\!&\!\!\!\! j_\infty(x)
=
|b(n)|^2
\rIm [\ov{A(x)}\na A(x)],
\\
\nonumber
\!\!\!\!\!\!\!\!\!\!\!\!\!\!\!\!\!\!\!\!\!\!\!\!&\!\!\!\!D\to\infty\!\!\!\!&\!\!\!\!
\eeqn
Here, we again  neglect the last sum of (\ref{laps}), for it does not contribute to the currents at large $|x|$.
\medskip\\
{\bf V.}
Finally, we
calculate the long range asymptotics of $A(x)$ as $|x|\to\infty$ and
show that
the convergence  (\ref{flu})
and  formula (\ref{ltast}) justify
(\ref{scs}),
(\ref{dics}) in the limit $D\to\infty$.

\medskip

 Let us comment on our methods.
We derive the limiting amplitude principle  (\ref{laps})  from
the dispersion decay in weighted energy norms by a~suitable
development of Agmon--Jensen--Kato's methods \cite{A,JK,KK2012}.
 The long range asymptotics (\ref{ecas}) is deduced from the `spherical version' (\ref{LSs})
of the Lippmann--\allowbreak Schwinger equation and a~refinement of  Lemma 3.2 from~\cite[Ch.~4]{BS}.
One of our key observations is that  {\it the spherical incident wave} from (\ref{ecas}) becomes asymptotically {\it the plane incident wave}
from (\ref{ecs}) as the source goes off to infinity: 
\begin{equation}\label{spw}
\fr{|q_D|}{|x-q_D|}e^{i|k|(|x-q_D|-|q_D|)}
\to e^{ikx}, \qquad D\to\infty.
\end{equation}
 In this limit, the picture of Fig.~\ref{Picture16} becomes the one of  Fig.~\ref{Picture15}.
We derive convergence (\ref{djR}) from
asymptotics (\ref{ecas})
by the Sobolev embedding theorem and the Ikebe uniqueness theorem
for the
Lippmann--\allowbreak Schwinger
equation \cite{Ik} (Theorem 3.1 of \cite[Ch.~4]{BS}).
Finally, we prove the second formula  of (\ref{jj}) for the flux (\ref{scfi})
in Theorem \ref{tM}. 
 We deduce it  from the decay of the oscillatory integrals (\ref{jouta3}), 
 which is due to  the interference of the incident and scattered waves.

\setcounter{theorem}{0}

\setcounter{equation}{0}
\section{Limiting amplitude principle}\label{LAP}

We  deduce the limiting amplitude principle  (\ref{laps}) from
the dispersion decay in weighted energy norms \cite{JK,KK2012}.
\bl\label{Llap}
Assume that conditions {\rm H0--H4} hold and $k\in\R^3$.
Then
\medskip\\
{\rm i)} The limiting amplitude principle  {\rm (\ref{laps})}
holds in the norm of
$\cH^2_{-\si}$ with any
$\si>5/2$.
\medskip\\
{\rm ii)} $C^l_q$ converge to the limit $C^l$
as $|q|\to\infty$,
and the limiting amplitude is given
by {\rm (\ref{las})}.
\el
\bpr
We should prove that
\begin{equation}\label{rho}
  \psi(x,t)= B_q(x)e^{-iE_k t}+\sum_1^NC^l_q\psi_l(x)e^{-iE^lt}
+r(x,t),
\end{equation}
where
\begin{equation}\label{Cr}
C^l_q\to C_l, \qquad |q|\to\infty; \qquad \Vert r(\cdot,t)\Vert_{\cH^2_{-\si}}\to 0, \qquad t\to\infty.
\end{equation}
The solution to the Cauchy problem (\ref{Scc}) is unique and is given by
the Duhamel representation
\begin{equation}\label{Duhr}
\psi(t)
=U(t)\psi^0-i\ds\int_0^t e^{-iE_k s}U(t-s)\rho_q~ds.
\end{equation}
Here, $U(t)$ is the dynamical group of equation (\ref{Scc}) with $\rho_q=0$,
and the first term in the right-hand side admits the expansion
\begin{equation}\label{fte0}
U(t)\psi^0=
\sum_1^NC^l\psi_le^{-iE^lt}
+r_0(t),
\end{equation}
where $C^l$ are independent of $q$, and
\begin{equation}\label{fte}
\Vert
r_0(t)
\Vert_{\cH^2_{-\si}}\le C\langle t\rangle^{-3/2}.
\end{equation}
This decay follows similarly to the dispersion decay in the norm $\cH^0_{-\si}$,
as established in (10.9) of~\cite{JK},  with suitable refinement of the resolvent high energy decay
(see Theorem 17.1 of \cite{KK2012}).
Here, the assumptions H0 and H3--H4 are essential.
\medskip\\
On the other hand,  the second term on the right-hand side of (\ref{Duhr}) can be written as
\begin{equation}\label{wed42}
I(t)=-i\ds\int_0^t e^{-iE_k s}U(t-s)\rho_q~ds=
-ie^{-iE_k t}\int_0^t e^{iE_k \tau}U(\tau)\rho_qd\tau.
\end{equation}
Here, $\rho_q\in \cH^2_{\si_1}$ with some $\si_1>5/2$
by H1. Hence, similarly to (\ref{fte0}) and (\ref{fte}),
\begin{equation}\label{fte1}
U(\tau)\rho_q=
\sum_1^ND^l_q\psi_le^{-iE^l \tau}
+r_q(\tau),
\end{equation}
where
\begin{equation}\label{fte2}
\Vert
r_q(\tau)
\Vert_{\cH^2_{-\si}}\le C_q\langle \tau\rangle^{-3/2}.
\end{equation}
Finally, the eigenfunctions $\psi_l(x)\in \cL^2_s$ with any $s\in\R$ by Agmon's theorem \cite[Theorem 3.3]{A},
(see also  Theorem 20.7 of \cite{KK2012}). Hence,
(\ref{Vcr}) implies that
\begin{equation}\label{clq}
D^l_q=\langle\rho_q, \psi_l\rangle=|q|\int\rho(x-q)\psi_l(x)dx=\cO(|q|^{-3-\ve_1}), \qquad |q|\to\infty.
\end{equation}
Therefore, $C^l_q=C^l+D^l_q\to C^l$ as $|q|\to\infty$, and
\begin{equation}\label{It}
I(t)\sim B_q(x)e^{-iE_k t}+\cO(|q|^{-3-\ve_1}), \qquad t\to\infty.
\end{equation}
Here, the asymptotics hold in $\cH^2_{-\si}$, and
the limiting amplitude is given by
\begin{equation}\label{Bqq}
B_q(x)=-i \int_0^\infty e^{iE_k \tau}U(\tau)\rho_qd\tau=
-i \int_0^\infty e^{i(E_k+i0) \tau}U(\tau)\rho_qd\tau,
\end{equation}
which can be written as (\ref{las}). This proves~(\ref{rho}).
\epr


\setcounter{theorem}{0}
\setcounter{equation}{0}

\section{Spherical waves}\label{SW}
In this section we obtain  long range asymptotics
(\ref{ecas}).
Denote $R=R(E_k+i0)$ and $R_0=R_0(E_k+i0)$,
where  $R_0(E)=(H_0-E)^{-1}$ is the resolvent of the free
Schr\"odinger operator $H_0=-\ds\fr12\De$.
Rewriting formula
(\ref{las}) for the limiting amplitude as the following `spherical version'
of the Lippmann--\allowbreak Schwinger equation, this gives
\begin{equation}\label{LSs}
  B_q(x)= R_0\rho_q(x)-R_0VB_q(x),
\end{equation}
since $R=R_0-R_0VR$. The free Schr\"odinger
resolvent $R_0(E)$
is the integral operator with kernel
$$
  R_0(E,x,y)=\fr{e^{i\sqrt{2E}|x-y|}}{2\pi|x-y|},\qquad
  E\in\C\setminus[0,\infty).
$$
Therefore, $R_0$ is the integral operator with kernel
\begin{equation}\label{R0E}
  R_0(E_k+i0,x,y)=\fr{e^{i|k||x-y|}}{2\pi|x-y|},
\end{equation}
because $\sqrt{2(E_k+i0)}=|k|$.
\medskip

For the first term on the right-hand side of
(\ref{LSs}),   asymptotics (\ref{ecas}) follow by a~suitable modification  of
Lemma 3.2 from  \cite[Ch.~4]{BS}. Let
\begin{equation}\label{S1}
S=\{\theta\in\R^3:|\theta|=1\}
\end{equation}
be the unit sphere.

\bl\label{lBS}
Under condition {\rm H1} with $\al=0$,
\begin{equation}\label{asBS}
  R_0\rho_q(x)=b\Big(\fr{x-q}{|x-q|}\Big)\fr{|q|}
  {1+|x-q|} e^{i|k| |x-q|}+K(x-q), \qquad x\in\R^3.
\end{equation}
Here, the amplitude $b\in C^1(S)$, and
\begin{equation}\label{bb}
  b(\theta)=\fr 1{2\pi}{\hat\rho(|k|\theta)}, \qquad |\theta|=1,
\end{equation}
where $\hat\rho(k)$ denotes the Fourier transform {\rm (\ref{Ft})}. The remainder
admits the bound
\begin{equation}\label{Kq}
|K(x-q)|
\le C|q|(1+|x-q|)^{-1-\ve_1}, \qquad x\in\R^3.
\end{equation}
\el
\bpr
This lemma follows
by the arguments from the proof of  Lemma 3.2 from \cite[Ch.~4]{BS},
with $|x|$ substituted almost everywhere by $|x|+1$.
Moreover, H1 with $\al=0$ implies that $\hat\rho\in   C^1_b(\R^3)$.
Hence, $b\in C^1(S)$ by (\ref{bb}).
\epr
As a~corollary, we obtain the bound
\begin{equation}\label{Rqhen}
  | R_0\rho_q(x)|\le \fr{C|q|}{1+|x-q|}, \qquad x\in\R^3.
\end{equation}

\begin{remark}
Our asymptotics \eqref{asBS} and the estimate \eqref{Kq}
differ from similar ones {\rm (3.50)} and {\rm (3.51)}  of {\rm \cite[Ch.~{\rm 4}]{BS}}, which hold only for $|x|\ge \de>0$.
\end{remark}

For the second term on the right-hand side of (\ref{LSs})
we need two additional technical lemmas.

\bl\label{lunib} Under conditions {\rm H1} and  {\rm H3}
the following bound holds for $k\ne 0${\rm :}
\begin{equation}\label{unib}
  \sup_{q\in\R^3}\Vert VB_q(x)\Vert_{\cL^2_\si}<\infty
  \quad \mbox{for any}\quad \si<5/2+\ve_2.
\end{equation}
\el
\bpr
The Lippmann--\allowbreak Schwinger equation (\ref{LSs}) implies
\begin{equation}\label{LSsi}
  (1+V R_0)VB_q= -VR_0\rho_q.
\end{equation}
On the other hand, $(1+V R_0)^{-1}=1-V R$. Hence,
\begin{equation}\label{LSso}
  VB_q= -(1-V R)VR_0\rho_q = -VR_0\rho_q +V RVR_0\rho_q.
\end{equation}
Let us
 estimate each term on the right-hand side separately.
\medskip\\
i) Condition (\ref{Vc}) with $\al=0$ and  bound
(\ref{Rqhen}) imply
\begin{equation}\label{Rqht}
   |VR_0\rho_q(x)|\le \fr {C|q|}{(1+|x-q|)(1+|x|)^{5+\ve_2}},\quad x\in\R^3.
\end{equation}
Therefore,
\begin{equation}\label{Rqhti}
  |VR_0\rho_q(x)|\le \fr {C}{(1+|x|)^{4+\ve_2}},\quad x\in\R^3.
\end{equation}
Hence,
\begin{equation}\label{vrr}
VR_0\rho_q\in \cL^2_{\si},\qquad \si<5/2+\ve_2.
\end{equation}
Thus, the bound (\ref{unib}) holds for the first term on the right-hand side of
(\ref{LSso}).
\medskip\\
ii) It remains to estimate the last term of (\ref{LSso}).
By (\ref{vrr}) we have $RVR_0\rho_q\in \cL^2_{-s}$ for any $s >1/2$, since the resolvent
$R=R(E_k+i0): \cL^2_s\to \cL^2_{-s}$ is continuous by \cite{JK}, Theorem 9.2,
because $E_k> 0$ for $k\ne 0$. Therefore, $VRVR_0\rho_q\in \cL^2_\si$ for
$\si <4.5+\ve_2$ by  (\ref{Vc}) with $\al=0$.
\epr

\bl Under conditions {\rm H1} and  {\rm H3}
the following uniform decay  holds{\rm :}
\begin{equation}\label{bah}
  \sup_{q\in\R^3}|R_0VB_q(x)|\le C(1+|x|)^{-2},\quad x\in\R^3.
\end{equation}
\el
\bpr
By (\ref{R0E}),  \begin{equation}\label{BSa}
  \begin{gathered}
  |R_0VB_q(x)|\le C\int\fr{|VB_q(y)|}{|x-y|}dy
 =
 C\int_0^\infty\Big [\int\fr{|VB_q(r,\vp,\theta)|d\vp\sin\theta
d\theta}
  {\sqrt{|x|^2+r^2-2|x|r\cos\theta}}\Big]r^2dr.
\end{gathered}
\end{equation}
Applying the Cauchy--Schwarz inequality to the inner integral,
\begin{multline}
|R_0VB_q(x)|
\le C\int_0^\infty
  \Big[\int |VB_q(r,\vp,\theta)|^2 d\vp\sin\theta d\theta\Big]^{\fr 12}
  \Big[\int\fr {d\vp\sin\theta d\theta}{|x|^2+r^2-2|x|r\cos\theta}
  \Big]^{\fr 12}r^2dr\\
=C\int_0^\infty\Big[\int|VB_q(r,\vp,\theta)|^2
  d\vp\sin\theta d\theta\Big]^{\fr 12}\Big[\fr{1}{|x|r}
  \log\fr{|x|+r}{||x|-r|}\Big]^{\fr 12}r^2dr.
 \label{BSo}
\end{multline}
Applying the same inequality to the last integral, this gives
\begin{gather} 
|R_0VB_q(x)| \nonumber\\
\le C\Big[\int(1+r)^{2\si}|VB_q(r,\vp,\theta)|^2
  d\vp\sin\theta d\theta r^2dr\Big]^{\fr 12}
\times \Big[\int_0^\infty
  \log\fr{|x|+r}{||x|-r|}~\fr{ r^2dr}{|x|r(1+r)^{2\si}}\Big]^{\fr 12}
  \nonumber\\
\le C(\si)\Big[\int_0^\infty\log\fr{|x|+r}{||x|-r|}~\fr{ rdr}
  {|x|(1+r)^{2\si}}\Big]^{\fr 12}
\label{int}
  =C(\si)\Big[\int_0^\infty
  \log\fr{1+ s}{|1- s|}~\fr{ |x| s d s}{(1+ s|x|)^{2\si}}
  \Big]^{\fr 12}
\end{gather}
for $\si<5/2+\ve_2$
by the uniform bound (\ref{unib}). Let us split the region of integration
$(0,\infty)=(0,1/2)\cup(1/2,3/2)\cup(3/2,\infty)$ and observe that
\begin{equation}\label{log}
\left.
\ba{ll}
\log\fr{1+ s}{|1- s|}=\cO(s),& \!\!\!\!\!\!\!\!\!\!\!\!\! s\in (0, 1/2)\\
\noalign{\smallskip}
\log\fr{1+ s}{|1- s|}\in L^1(1/2,3/2)& \\
\noalign{\smallskip}
\log\fr{1+ s}{|1- s|}=\cO(s^{-1}),& \!\!\!\!\!\!\!\!\!\!\!\! s\in (3/2,\infty)
\ea
\right|
\end{equation}
Then the integral (\ref{int}) can be  estimated as
\beqn\label{lain}
&&C_1(\si)\Big[
\int_0^{1/2}\fr{ |x| s^2 d s}{(1+ s|x|)^{2\si}}
  +(1+|x|)^{1-2\si}+\int_{3/2}^\infty\fr{ |x| d s}{(1+ s|x|)^{2\si}} \Big]
  \nonumber \\
  \nonumber\\
  \nonumber
&&=C_1(\si)\Big[
|x|^{-2}\int_0^{|x|/2}\fr{ r^2 dr}{(1+r)^{2\si}}+(1+|x|)^{1-2\si}
  +\int_{3|x|/2}^\infty\fr{ dr}{(1+r)^{2\si}}\Big]\\
  \nonumber\\
&&\le C_2(\si)
(1+|x|)^{1-2\si}
\eeqn
for $2\si>1$.
This gives (\ref{bah}), since we can take any $\si<5/2+\ve_2$ by (\ref{unib}).
\epr
Now we are ready to prove (\ref{ecas}).

\bp\label{pbla}
Asymptotics {\rm (\ref{ecas})} hold
under conditions   {\rm H1--H3}.
\ep
\bpr
The Lippmann--\allowbreak Schwinger equation (\ref{LSs}) yields
$$
VB_q(x)= -VR_0\rho_q(x)-VR_0VB_q(x).
$$
Hence, (\ref{Vc}) with $\al=0$ and (\ref{Rqhti}), (\ref{bah}) imply that
\begin{equation}\label{coi}
  |VB_q(x)|\le \fr C{(1+|x|)^{4+\ve_2}}.
\end{equation}
Therefore, similarly to  (\ref{asBS}), we obtain  the asymptotics
\begin{equation}\label{asBS2}
  R_0VB_q(x)=c_q\Big(\fr{x}{|x|}\Big)
\fr{ e^{i|k| |x|}}{1+|x|}+L_q(x),
\end{equation}
where
\begin{equation}\label{Kq22}
  |L_q(x)|\le C(1+|x|)^{-1-\ve_2}.
\end{equation}
Now (\ref{LSs}) and  (\ref{asBS}), (\ref{asBS2}) imply
\begin{equation}\label{ecas0}
  B_q(x)\sim b\Big(\fr{x-q}{|x-q|}\Big)\fr{|q|}
  {|x-q|} e^{i|k| |x-q|}+c_q\Big(\fr{x}{|x|}\Big)\fr{e^{i |k|\cdot|x|}}{|x|}
  \end{equation}
as $ |x-q|\to\infty$ and $|x|\to\infty$.
Denote $B_D(x):=B_{q_D}(x)$, where $q_D=-nD$ with $n=k/|k|$ and $D>0$. Then
\begin{equation}\label{ecas1}
B_D(x)\sim b(n)\Big[\fr{|q_D|}{|x-q_D|} e^{i|k| |x-q_D|}
  +d_D(k,\theta)\fr{e^{i |k|\cdot|x|}}{|x|}\Big]
\end{equation}
as $|x-q_D|\to\infty$ and $|x|\to\infty$,
where $\theta:=x/|x|$, because $b(n)\sim \hat\rho(|k|n)\ne 0$ by (\ref{bb}) and
the Wiener condition H2. This is the only point in our analysis,
where the  Wiener condition is called for. Finally,  (\ref{ecas1}) can be written as
(\ref{ecas}) with $b_D(n):=b(n)e^{i|k|D}$ and
$\ds a_D(k,\theta)=d_D(k,\theta)e^{-i|k|D}$.
\epr

The following corollary is of crucial importance in the next section.

\bc
Bound  {\rm (\ref{Rqhen})}, asymptotics  \eqref{asBS2}--\eqref{Kq22}, and
 formula  {\rm (\ref{LSs})}
imply
\begin{equation}\label{unibo}
  | B_q(x)|\le \fr{C|q|}{1+|x-q|}+\fr C{(1+|x|)}, \qquad x,q\in\R^3.
\end{equation}
\ec

\setcounter{theorem}{0}
\setcounter{equation}{0}

\section{Plane wave limit}\label{PWL}
In this section we prove convergence (\ref{djR}) from  the uniqueness of solution to the Lipp\-mann--\allowbreak Schwin\-ger equation
\begin{equation}\label{LS}
  A(x)=e^{ik x} -R_0VA(x),
\end{equation}
which is equivalent to (\ref{ltast}) (see Lemma \ref{lTa} below).
First, we rewrite (\ref{LSs}) with $q=q_D$ as
\begin{equation}\label{LSsh}
  A_D(x)= R_0\rho_{q_D}(x)/b_D(n)-R_0VA_D(x),
\end{equation}
where $A_D(x):=B_D(x)/b_D(n)$.
By (\ref{asBS}) and (\ref{spw}) the first term on the right-hand side
of (\ref{LSsh}) converges to the first term on the right-hand side of (\ref{LS}),
\begin{equation}\label{con1}
  R_0\rho_{q_D}(x)/b_D(n)\to e^{ikx}, \qquad D\to\infty,
\end{equation}
in  $C(\R^3)$.
Now
(\ref{djR}) means the convergence
of  the corresponding
solutions:

\bp\label{pAD}
Let conditions  {\rm H1--H3}  hold and let $k\ne 0$. Then the convergence
\begin{equation}\label{djR1}
  A_D(x)\to A(x), \qquad D\to\infty
\end{equation}
holds in $\cH^s_{-\si}$ with any $s<2$ and $\si>5/2$, where
the function $A(x)$ is defined by {\rm (\ref{ltast})}.

\ep
\bpr
We  deduce the convergence from the compactness of the family
$\{A_D(x): D>0\}$ and Ikebe's uniqueness theorem  \cite{Ik} (Theorem 3.1 of \cite[Ch.~4]{BS}).
\medskip\\
{\it Step} i).  By (\ref{LSsh}),
$$
  \Vert A_D\Vert_{\cH^2_{-\si}}
  \le C\Vert R_0\rho_{q_D}\Vert_{\cH^2_{-\si}}
  +\Vert R_0VB_D\Vert_{\cH^2_{-\si}}.
$$
The first term on the right-hand side is uniformly bounded for
$D>0$, since estimate of type (\ref{Rqhti}) holds
with $\langle x\rangle^{-\si}$ and $\si$
instead of $V(x)$ and $5+\ve_2$, respectively.
The second term is uniformly  bounded, since $VB_q$ is uniformly bounded
in $\cL^2_\si$ with $\si<5/2+\ve_2$ by (\ref{unib}), while the operator
$R_0: \cL^2_s\to \cH^2_{-s}$ is continuous for any $s>1/2$
by   Theorem 18.3 i) of \cite{KK2012}, because  $k\ne 0$.
Hence,
\begin{equation}\label{unib2}
  \sup_{D>0}\Vert A_D\Vert_{\cH^2_{-\si}}<\infty, \qquad \si >5/2.
\end{equation}
{\it Step} ii).
Now  the Sobolev embedding theorem  \cite{LM} implies
that the family  $\{A_D(x): D>0\}$ is a~precompact set in the Hilbert space
$\cH^s_{-\si}$ with any $s<2$ and $\si >5/2$.
Hence, for any sequence $D_j\to\infty$, there is a~subsequence $D_{j'}\to\infty$ such that
\begin{equation}\label{coms}
A_{D_{j'}}(x)\to A_*(x), \qquad j'\to\infty,
\end{equation}
where the convergence holds in $\cH^s_{-\si}$ with any $s<2$
 and $\si>5/2$.
Therefore,
\begin{equation}\label{coms2}
VA_{D_{j'}}(x)\to VA_*(x), \qquad j'\to\infty,
\end{equation}
where the convergence holds in $\cH^s_\si$ with $s<2$ and
some
$\si>5/2$ by H3.
\medskip\\
{\it Step} iii). At last,  equation (\ref{LSsh}) and  convergences (\ref{coms}),
(\ref{coms2}), and (\ref{con1}) imply equation (\ref{LS}) for $A_*(x)$:
\begin{equation}\label{LS*}
A_*(x)=e^{ik x} -R_0VA_*(x),
 \end{equation}
since the operator $R_0:=R_0(E_k+i0):\cL^2_\si\to \cL^2_{-\si}$ is continuous
for $\si>1/2$ by Lemma 2.1 of \cite{JK}.

The function $A_*(x)$ is bounded by  (\ref{unibo}) and is continuous
by the Sobolev embedding theorem, since  $A_*(x)\in \cH^s_{-\si}$
with any $s<2$ and $\si>5/2$ by  (\ref{coms}).

Finally, $A(x)=A_*(x)$ by Ikebe's uniqueness theorem
\cite{BS,Ik},  which holds for $k\ne 0$
under the condition (\ref{Vcr})
for  bounded continuous solutions to the Lippmann--\allowbreak Schwinger equation
(\ref{LS}). Hence,
 convergence (\ref{coms})
implies (\ref{djR1}), since the limit function $A_*(x)$
does not depend on the subsequence~$j'$.
\epr

\begin{remark}
{\rm Let us emphasize that the right-hand side of (\ref{unibo})
with  $q=-nD$ is not uniformly bounded for $D>0$: its value at $x=q$ tends to infinity
as $D\to\infty$. Nevertheless, (\ref{unibo}) implies that every limit function $A_*(x)$ is bounded.

}
\end{remark}


\setcounter{equation}{0}
 \setcounter{theorem}{0}
\section{Convergence of flux}\label{CF}

We check the convergence of the limit flux as the source goes off to infinity.
First, we use (\ref{laps}) and (\ref{djR}) to verify (\ref{laps2}) and (\ref{flu}).

\bl\label{pfD}
Under conditions {\rm H1--H3}
the convergence
\begin{equation}\label{laps22}
  \vp_D(x,t)/b_D(n)\to A(x)e^{-iE_kt}, \qquad D\to\infty, \quad t\in\R,
\end{equation}
holds in  $\cH^s_{-\si}$ with any $s<2$ and $\si>5/2$.
\el
\bpr
The convergence follows from Proposition \ref{pAD}, since
\begin{equation}\label{laps23}
\vp_D(x,t)/b_D(n)=
A_D(x)e^{-iE_k t}
\end{equation}
as $t\to\infty$
by the definition of $\vp_q(x,t)$ in (\ref{laps}) with $q(D)=-nD$
(recall that we omit the sum over the discrete spectrum
in  (\ref{laps})).
Here, $|b_D(n)|=|b(n)|\ne 0$ by the Wiener condition
 (\ref{Wr}).
\epr
\bc \label{cflu}
{\rm i)} Convergence {\rm (\ref{flu})} holds in $\cL^2_{\rm loc}(\R^3)$:
\begin{equation}\label{fluc}
  j_D(x)=\rIm [[\ov{\vp_D(x,t)}\na \vp_D(x,t)]
  \longrightarrow  j_\infty(x)=
  |b(n)|^2\rIm [\ov{A(x)}\na A(x)], \qquad D\to\infty.
\end{equation}
{\rm ii)}
Moreover, the convergence holds `in the sense of flux'; i.e.,
\begin{equation}\label{cjs}
  \int_S j_D(x)\cdot \nu(x) dS(x)\to\int_S
  j_\infty(x)\cdot \nu(x) dS(x), \qquad D\to\infty, \quad t\in\R,
\end{equation}
for any  compact smooth two-dimensional submanifold $S\subset\R^3$
with boundary, where $\nu(x)$ is  the unit normal field
to $S$ and $dS(x)$ stands for the corresponding Lebesgue measure on $S$.
\ec
\bpr
The convergence (\ref{laps22}) also holds in $C(\R^3)$, since  $\cH^s_{-\si}\subset C(\R^3)$
for $s>3/2$ by the Sobolev embedding theorem \cite{LM}.
Further, the convergence of the derivatives
\begin{equation}\label{laps22d}
  \na\vp_D(x,t)/b_D(n)\to \na A(x)e^{-iE_kt},\qquad D\to\infty, \quad t\in\R,
\end{equation}
holds in $\cH^{s-1}_\si$ with any $s<2$. Hence, the
convergence (\ref{fluc}) holds in $\cL^1_{\rm loc}(\R^3)$, and
moreover,
\begin{equation}\label{laps22dt}
  \na\vp_D(x,t)/b_D(n)\Big|_S\to \na A(x)e^{-iE_kt}\Big|_S,
  \qquad D\to\infty, \quad t\in\R,
\end{equation}
in $\cL^2(S)$ by the Sobolev trace theorem \cite{LM}, for we can take $s>3/2$.

Similarly, (\ref{laps22}) also implies the convergence in  $\cL^2(S)$
\begin{equation}\label{laps22dt2}
  \vp_D(x,t)/b_D(n)\Big|_S\to  A(x)e^{-iE_kt}\Big|_S,
  \qquad D\to\infty, \quad t\in\R,
\end{equation}
Therefore, the integrands in (\ref{cjs}) converge in  $\cL^1(S)$, inasmuch as $|b_D(n)|=|b(n)|\ne 0$.\epr


\setcounter{equation}{0}
 \setcounter{theorem}{0}

\section{Long range asymptotics}
We obtain asymptotics (\ref{ecs}).
The first lemma is well known \cite{Ta}.

\bl\label{lTa}
Equation {\rm (\ref{LS})} admits a~unique bounded continuous solution, which is
 given by {\rm (\ref{ltast}):}
\begin{equation}\label{LSu}
  A(x)= e^{ik x} -RVe^{ik x}
\end{equation}
\el
\bpr
We should prove  (\ref{LSu}) assuming (\ref{LS}). First, we apply
the general operator identity
$$
P^{-1}=Q^{-1}+Q^{-1}(Q-P)P^{-1}
$$
 to
$P=H_0-E_k-i0$ and $Q=H-E_k-i0$. Then we obtain $R_0=R+RVR_0$,
and hence
$$
  R_0VA=RVA+RVR_0VA=RV(A+R_0VA)=RVe^{ik x}
$$
by (\ref{LS}). Substituting into (\ref{LS}), we obtain (\ref{LSu}).
\epr

Next, we need an extension of Lemma \ref{lBS} to functions from weighted Ag\-mon--So\-bolev spaces.

\bl\label{lBSg}
Let   $r(x)\in \cH^2_\si$ for some $\si>7/2$. Then
\begin{equation}\label{asBSg}
  R_0r(x)=\phi(\theta)\fr {e^{i|k| |x|}}{|x|} +K(x), \qquad \theta:=\fr{x}{|x|},\qquad |x|>1.
\end{equation}
Here, the amplitude
$\phi\in C^2(S)$, and
\begin{equation}\label{bbg}
  \phi(\theta)=\fr 1{2\pi}{\hat r(|k|\theta)},\quad |\theta|=1.
\end{equation}
The remainder admits the bounds
\begin{equation}\label{Kqg}
  |K(x)|\le C|x|^{-2},~~~|\na K(x)|\le C|x|^{-2},~~~|\na\na K(x)|\le C|x|^{-2}, \qquad |x|>1.
\end{equation}
\el
\bpr
First,
$$
  R_0r(x)=\fr{e^{i|k||x|}}{2\pi|x|}\int e^{-i|k|\fr x{|x|}\cdot y}r(y)dy+
  \fr 1{|x|}\int \fr{\langle y\rangle^2}{|x-y|}R(x,y) r(y)dy,
$$
where the function $R(x,y)$ is bounded as in the proof of Lemma 3.2 from \cite[Ch.~4]{BS}. Hence,
formula (\ref{bbg}) follows with $\phi\in C^2(S)$, since $\hat r\in \cH^\si_2\subset C^2_b(\R^3)$ for $\si>7/2$
by the Sobolev embedding theorem.

To prove the first estimate of (\ref{Kqg}), it suffices to check that
$$
  J(x):=\int \fr{\langle y\rangle^2}{|x-y|} |r(y)|dy\le C|x|^{-1},\qquad |x|>1.
$$
Using the Cauchy--Schwarz inequality, we obtain
$$
  |J(x)|\le \Big(\int \fr{1}{|x-y|^2\langle y\rangle^{2\si-4}}dy\Big)^{1/2}
  \Vert r\Vert_{\cL^2_\si}.
$$
Now it suffices to prove the bound
\begin{equation}\label{I}
  I(x):=\int \fr{1}{|x-y|^2\langle y\rangle^{2\si-4}}dy\le C|x|^{-2},\qquad |x|>1.
\end{equation}
In the spherical coordinates,  we obtain similarly to (\ref{int})--(\ref{lain}),
\begin{align*}
  I(x)&=2\pi\int_0^\infty\fr{r^2dr}{(1+r)^{2\si-4}}
  \int_0^\pi \fr{\sin\theta d\theta}{|x|^2+r^2-2|x|r\cos\theta}\\
 &=2\pi|x|\int_0^\infty\fr{s ds}{(1+s|x|)^{2\si-4}}\log
  \fr{|1+s|}{|1-s|}\\
  &\le C\int_0^{1/2}\fr{ |x|s^2 ds}{(1+s|x|)^{2\si-4}}
  +C|x|^{5-2\si}+C\int_{3/2}^\infty\fr{ |x| ds}{(1+s|x|)^{2\si-4}}\\
  &=C|x|^{-2}\int_0^{|x|/2}\fr{ r^2 dr}{(1+r)^{2\si-4}}
  +C|x|^{5-2\si}+C\int_{3|x|/2}^\infty\fr{ dr}{(1+r)^{2\si-4}}\\
&\le C_1(\si)|x|^{-2}+C_2|x|^{5-2\si} \le C|x|^{-2},\qquad |x|>1,
\end{align*}
since $2\si>7$.
This proves the first bound in~(\ref{Kqg}).

To prove the second bound in (\ref{Kqg}), we differentiate (\ref{asBSg}):
\begin{equation}\label{asBSgd}
  \na R_0r(x)=\phi(\theta)i|k|\theta\fr {e^{i|k| |x|}}{|x|}
  + \cO(|x|^{-2}) +\na K(x),\qquad |x|\to\infty.
\end{equation}
On the other hand, $\na R_0r(x)= R_0\na r(x)$, where $\na r(x)\in\cH^1_\si$.
Hence, by the above arguments,
\begin{equation}\label{asBSa}
  \na R_0r(x)=\phi_1(\theta)\fr{e^{i|k| |x|}}{|x|}+ \cO(|x|^{-2}),
  \qquad |x|\to\infty,
\end{equation}
where
$$
  \phi_1(\theta)=\fr 1{2\pi}\widehat{\na r}(|k|\theta)
  =\fr 1{2\pi}i|k|\theta\widehat{r}(|k|\theta)=
  i|k|\theta\phi(\theta).
$$
So, the second bound in (\ref{Kqg}) follows by comparing  (\ref{asBSgd}) and (\ref{asBSa}).

The last bound of (\ref{Kqg}) follows similarly.
\epr

Now asymptotics of type (\ref{ecs}) follow from (\ref{LSu}) and
the next
lemma, which is a~refinement of Theorem 3.2 from \cite[Ch.~4]{BS}.

\bl \label{l7}
Let  condition {\rm H3} hold and let $k\ne 0$. Then
\begin{equation}\label{idea}
  -RVe^{ikx}(x)=a(k,\theta)\fr{e^{i|k||x|}}{|x|}+K_1(x), \qquad \theta:=\fr{x}{|x|}, \qquad |x|>1.
\end{equation}
The amplitude  $a(k,\cdot)\in C^2(S)$ is given by
{\rm (\ref{aT})},
and the remainder admits the bound
\begin{equation}\label{Kqg1}
  |K_1(x)|+|\na K_1(x)|+|\na\na K_1(x)|\le C|x|^{-2},\qquad |x|>1.
\end{equation}
\el
\bpr
First, $RV=R_0 T$,
where $T:=T(E_k+i0)$
(see (3.31) of \cite[Ch.~4]{BS}, and \cite{Ta}). Hence,
\begin{equation}\label{ideah}
  -[RVe^{ikx}](x)=-[R_0Te^{ikx}](x).
\end{equation}
Therefore, (\ref{idea})--(\ref{Kqg1}) will follow from Lemma \ref{lBSg} if we verify that
 \begin{equation}\label{7}
 Te^{ikx}\in \cH^2_\si, \qquad \forall \si<7/2+\ve_2/2.
 \end{equation}
 Indeed,
$$
Te^{ikx}=Ve^{ikx}-VRVe^{ikx},
$$
 where $Ve^{ikx}\in  \cH^2_\si$ with any $\si< 7/2+\ve_2/2$
by H3. Hence, $RVe^{ikx}\in  \cH^2_s$ with any $s<-1/2$
by Corollary 19.3 of \cite{KK2012}, since $k\ne 0$.
Therefore,
 $VRVe^{ikx}\in  \cH^2_\si$ with
any $\si<9/2+\ve_2$ by H3.

Finally, applying Lemma \ref{lBSg} to the function $r(x)=Te^{ikx}$
and using (\ref{ideah}), we obtain asymptotics (\ref{idea})--(\ref{Kqg1}) with the amplitude given by (\ref{aT}):
\begin{equation}\label{ate}
  a(k,\theta)=-\fr 1{2\pi}\hat r(|k|\theta)=-\fr 1{2\pi}(Te^{ikx},e^{i|k|\theta x})=
  -4\pi^2T(|k|\theta,k)
\end{equation}
according to (\ref{Tm}).
\epr
\begin{remark}
{\rm Formula  (\ref{aT}) for
the amplitude in (\ref{idea}) is well known, see formula   (97a) of
 \cite{RS}.  On the other hand, the asymptotics
 (\ref{Kqg1}) and the fact that $a(k,\cdot)\in C^2(S)$
 are new, to our knowledge, and
  play the key role in the next section.}
\end{remark}

\setcounter{equation}{0}
 \setcounter{theorem}{0}
\section{Differential cross section}
\label{LRA}

Now we can justify formula (\ref{scsa}). Convergence (\ref{laps22}) and the formula (\ref{LSu}) imply
the asymptotics of the limiting amplitudes,
\begin{equation}\label{vpD}
  \vp_D(x,t)/b_D(n)\!\sim\!f(x,t)\!:=\! A(x)e^{-iE_kt}\!=\!e^{ikx}e^{-iE_kt}
  \!-RV[e^{ikx}]e^{-iE_kt},~~D\to \infty,
\end{equation}
which holds in $\cH^s_{-\si}$ with any $s<2$ and $\si>5/2$.

Let us recall that  $j_D(x,t)$ and  $j_\infty(x,t)$ denote the
total flux (\ref{Sflu}), corresponding to the wave fields
$\vp_D(x,t)$ and
$b(n)f(x,t)$, respectively (see (\ref{fluc})).
The convergence (\ref{cjs})
 means that the limiting current $j_\infty(x,t)$
can be `measured' in the double limit: first, as $t\to\infty$, and then, as $D\to\infty$.

The corresponding scattered flux should be defined as the difference (\ref{scfi}):
\begin{equation}\label{scf}
j^{\rm sc}(x,t)=j_\infty(x,t)-j^{\rm in},
\end{equation}
where $j^{\rm in}:=\lim_{|x|\to\infty} j_\infty(x,t)$.
Let us adjust the meaning of the angular density of the scattered flux (\ref{jout}).

\bd
The limit {\rm (\ref{jout})} means that
\begin{equation}\label{jouta}
R^2\int_S \phi(\theta)
j^{\rm sc}(x,t)
\cdot\theta ~d\theta\to \int_S \phi(\theta)j_a^{\rm sc}(\theta) d\theta, \qquad R\to\infty
\end{equation}
for any test function $\phi\in C^\infty(S)$ with $\phi(\theta)=0$ in a~neighborhood of $\theta=\pm n$.
\ed
In other words, the limit (\ref{jout}) is understood in the sense of distributions on $S\setminus\{n,-n\}$.

\bt\label{tM} Under assumption H3,
\begin{equation}\label{M}
  j^{\rm in}= |b(n)|^2k, \qquad
  j_a^{\rm sc}(\theta) =  |b(n)|^2|a(k,\theta)|^2|k|.
\end{equation}
\et
\bpr
Using (\ref{Sflu}) and (\ref{vpD}),
\begin{equation}\label{flux}
  j_\infty(x,t)= |b(n)|^2\rIm [\ov{f(x,t)}\na f(x,t)]=
  |b(n)|^2\rIm [(\ov{e^{ikx}+a^{\rm sc}(x)) }\na (e^{ikx}+a^{\rm sc}(x) )].
\end{equation}
Here, the amplitude $a^{\rm sc}(x)=-RV[e^{ikx}]$ decays at infinity together with its
derivatives
according to (\ref{idea})--(\ref{Kqg1}). Hence, the flux (\ref{flu}) for large $|x|$ equals $|b(n)|^2k$,
which proves the first formula of (\ref{M}).

It remains to prove the second formula of (\ref{M}). According to definition (\ref{jouta}),
we should check that
\begin{equation}\label{jouta2}
\begin{gathered}
R^2\int_S \phi(\theta)
\rIm [e^{-ikR\theta} \na a^{\rm sc}(R\theta) +\ov{a^{\rm sc}(R\theta)}ik e^{ikR\theta}+ \ov{a^{\rm sc}(R\theta)}\na a^{\rm sc}(R\theta)] \cdot\theta ~d\theta
\\
\to \int_S \phi(\theta) |a(k,\theta)|^2|k| d\theta, \qquad R\to\infty.
\end{gathered}
\end{equation}
Here, \begin{equation}\label{aa}
\ov{a^{\rm sc}(R\theta)}\na a^{\rm sc}(R\theta)\cdot\theta=|a(k,\theta)|^2|k|R^{-2}+\cO(R^{-3})
\end{equation}
by Lemma \ref{l7}.
Hence, it remains to prove that the oscillatory integrals in (\ref{jouta2}) vanish in the limit as $R\to\infty$.
This follows by the partial integration
in view of Lemma \ref{l7}, since the phase functions do not have stationary points outside $\theta=\pm n$.
Indeed, let us consider, for example, the  oscillatory integral
\begin{equation}\label{jouta3}
\begin{gathered}
R^2\int_S \phi(\theta)
e^{-ikR\theta} \na a^{\rm sc}(R\theta)  \cdot\theta ~d\theta=
R^2\int_S \phi(\theta)
e^{-ikR\theta} \na [ a(k,\theta)\fr{e^{i|k|R}}{R}+K_1(R\theta)]  \cdot\theta ~d\theta \\
=
R^2\int_S \phi(\theta)
e^{-ikR\theta} [ a(k,\theta)\fr{i|k|\theta e^{i|k|R}}{R}-a(k,\theta)\fr{e^{i|k|R}}{R^2}\theta+ \na a(k,\theta)\fr{e^{i|k|R}}{R}+
\na K_1(R\theta)]  \cdot\theta ~d\theta.
\end{gathered}
\end{equation}
Here, the phase functions $kR\theta$ and $kR\theta-|k|R$ admit exactly two stationary points $\theta=\pm n=\pm k/|k|$ on the
sphere $S$.
Hence,  the decay for each integral
in the last line of
(\ref{jouta3}) follows  by the partial integration.
The integrals
with $a(k,\theta)$ vanish in the limit $R\to\infty$, since $a(k,\cdot)\in C^2(\R^3)$:
the first  integral
vanishes  by twofold
partial integration, while the second and the third ones, by the single partial integration.
The integral with $\na K_1$ vanishes in the limit $R\to\infty$ by
the single partial integration due to (\ref{Kqg1}).
\epr

\bc\label{cM}
According  to {\rm (\ref{vpD})} and {\rm (\ref{M})}, the differential cross section in the limit $D\to\infty$
is given by
$$
  \si(\theta):=j_a^{\rm sc}(\theta)/|j^{\rm in}|=|a(k,\theta)|^2,
$$
which justifies {\rm (\ref{scsa})}. Then {\rm (\ref{dics})} also holds by the known formula {\rm (\ref{aT})}.
\ec




\begin{thebibliography}{}

\bibitem{A}
Agmon S., Spectral properties of Schr\"odinger operator and scattering theory,
{\em Ann.~Scuola Norm.~Sup.~Pisa}, Ser.~IV {\bf 2}, 151--218 (1975).

\bibitem{AJS}
W.~Amrein, J.~Jauch, K.~Sinha,
Scattering Theory in Quantum Mechanics.
Physical Principles and Mathematical Methods,
 W. A. Benjamin, London, 1977.



\bibitem{BS}
F.A. Berezin, M.A. Shubin,
The Schr\"odinger Equation,
Kluwer Academic Publishers, Dordrecht, 1991.



\bibitem{Bor}
V.A.~Borovikov,
 Diffraction by Polygons and Polyhedrons,
Nauka, Moscow, 1966.

 \bibitem{BSp}
M. Butz, H. Spohn,
 Dynamical phase transition for a~quantum particle source,
 {\em Ann. Henri Poincar\'e} {\bf 10} (2010), no. 7, 1223--1249.
 arXiv:0908.2912


\bibitem{CNS}
J.M.~Combes, R.G.~Newton, R.~Stokhamer, {\em Phys. Rev. D} {\bf 11} (1975), no. 2, 366--372.

\bibitem{D69}
J.D.~Dollard, Scattering into cones I: Potential scattering,
{\em Commun. Math. Phys.} {\bf 12} (1969), no.~3, 193--203.

\bibitem{DGMZ}
D.~D\"urr, S.~Goldstein, T.~Moser, N.~Zangh\`i,
A microscopic derivation of the quantum mechanical
formal scattering cross section, {\em Commun. Math. Phys.} {\bf 266} (2006), no.~3, 665--697.

\bibitem{DGTZ}
D.~D\"urr, S.~Goldstein, S.~Teufel, N.~Zangh\`i,
Scattering theory from microscopic first principles, {\em Physica A} {\bf 279} (2000), no.~1--4, 416--431.

\bibitem{DT}
D.~D\"urr,  S.~Teufel,
Bohmian Mechanics.
The Physics and Mathematics of Quantum Theory,
Springer, 2009.

\bibitem{DMP}
D.~D\"urr,
T.~Moser, P.~Pickl,
The flux-across-surfaces theorem under
conditions on the scattering state,
{\em J. Phys. A, Math. Gen.} {\bf 39} (2006), no.~1, 163--183.

\bibitem{Eidus}
D.M. Eidus,
The limiting amplitude principle for the Schr\"odinger
equation in domains with unbounded boundaries,
{\em Asymptotic Anal.} {\bf 2} (1989), No.~2, 95--99.

\bibitem{ES}
V.~Enss, B.~Simon,
Finite total cross-sections in nonrelativistic quantum mechanics,
{\em Commun. Math. Phys.} {\bf 76}  (1980), 177--209.

 \bibitem{Ik}
T.~Ikebe,
Eigenfunction expansions associated with
the Schr\"odinger operators and their applications
to scattering theory, {\em Arch. Ration. Mech. Anal.} {\bf 5} (1960), 1--34.

\bibitem{JLN}
J.M.~Jauch, R.~Lavin, R.G.~Newton,
Scattering into cones, {\em Helv. Phys. Acta} {\bf 45} (1972), 325--330.

\bibitem{JK}
Jensen A., Kato T., Spectral properties of Schr\"odinger operators
and time-decay
of the wave functions, {\em Duke Math.~J.}{\bf\,\,46}, 583--611 (1979).

 \bibitem{KK2012}
 A. Komech, E. Kopylova,
 Dispersion Decay and Scattering Theory,  John Wiley \& Sons, Hoboken, NJ, 2012.

\bibitem{LM}
J.L. Lions, E. Magenes,
Non-homogeneous boundary value problems and applications,
Vol.~I,  Springer, Berlin, 1972.

\bibitem{M}
Murata M., Asymptotic expansions in time for solutions of
Schr\"odinger-type equations, {\em J.~Funct.~Anal.}{\bf\,\,49}, 10--56 (1982).

\bibitem{New}
 R.G.~Newton, Scattering Theory of Waves and Particles, Springer, NY, 1982.

 \bibitem{Po}
A.Ya. Povzner, On the expansion of arbitrary functions
in characteristic functions of the operator $-\Delta u+cu$,
{\em Mat. Sbornik N.S.} {\bf   32(74)} (1953), 109--156. [Russian]

\bibitem{RS}
Reed M., Simon B., Methods of Modern Mathematical Physics, III:
Theory of Scattering, Academic Press, 1979.

\bibitem{Sakurai}
J.J. Sakurai, Advanced Quantum Mechanics, Addison-Wesley, Massachusetts, 1967.

\bibitem{Schiff}
L.I.~Schiff, Quantum Mechanics, McGraw-Hill, NY, 1968.

\bibitem{Ta}
J.R.~Taylor, Scattering Theory, Wiley, NY, 1972.

\bibitem{TDM}
S.~Teufel, D.~D\"urr, K.~M\"unch-Berndl, The flux-across-surfaces theorem for
short range potentials and wave functions without energy cutoffs,
{\em J. Math. Phys.} {\bf  40} (1999), no.~4, 1901--1922.

\bibitem{T}
S.~Teufel, The flux-across-surfaces theorem and its implications
on scattering theory, Dissertation,
M\"unchen: Univ. M\"unchen, Fakult\"at f\"ur Mathematik und Informatik, 1999. http://www-m5.ma.tum.de/pers/teufel/Diss.ps


\bibitem{Weinberg}
 S.~Weinberg, The Quantum Theory of Fields. Vol.~1. Foundations,
 Cambridge University Press, Cambridge, 2005.

 \bibitem{Yaf}
D.R. Yafaev, Mathematical Scattering Theory: Analytic Theory, AMS, Providence, Rhode Island, 2010.
\end{thebibliography}
\end{document}